\documentstyle[11pt]{article}
\begin{document}

\title{Integrable Schr\"odinger operators with magnetic fields:
factorisation method on curved surfaces}
\author{{\Large E.V. Ferapontov  and A.P. Veselov } \\
    Department of Mathematical Sciences \\ 
    Loughborough University \\
    Loughborough, Leicestershire LE11 3TU, U.K. \\
     and \\
    Landau Institute for Theoretical Physics\\
    Academy of Science of Russia, Kosygina 2\\
    117940 Moscow, Russia\\
    e-mails:\\
   {\tt E.V.Ferapontov@lboro.ac.uk} \\
    {\tt A.P.Veselov@lboro.ac.uk} \\
  }
\date{}
\maketitle

\newtheorem{theorem}{Theorem}

\pagestyle{plain}

\maketitle

\begin{abstract}
The factorisation method for Schr\"odinger operators with magnetic fields on
a two-dimensional surface $M^2$ with non-trivial metric is investigated.
This leads to the new integrable examples of such operators and brings
a new look at some classical problems such as Dirac magnetic monopole and Landau problem.
The global geometric aspects and related spectral properties 
of the operators from the factorisation chains are discussed in details.
We also consider the Laplace transformations on a curved surface and extend 
the class of Schr\"odinger operators with two integrable levels
introduced in the flat case by S.P.Novikov and one of the authors. 
\end{abstract}

\section{Introduction}
In spite of the fact that some important examples of integrable Schr\"odinger equations with magnetic 
fields were known since 1930-th 
(Landau problem, Dirac magnetic monopole), the general problem of integrability for such equations is still 
far from being understood. 

In two dimensions probably the first important step in this direction has been done
by Dubrovin, Krichever and Novikov in 1976, who introduced a very important class of Schr\"odinger operators 
with magnetic fields integrable (``finite-gap") on one energy level \cite{DKN76}. The coefficients of the 
corresponding operators are periodic (or quasiperiodic) so that the total magnetic flux is zero.

The case of periodic magnetic fields with non-zero flux has been considered by 
Dubrovin and Novikov \cite{DN80} following Aharonov-Casher observation \cite{AC} that the Pauli operators for spin $1/2$ particles
in magnetic field are related to the factorisable Schr\"odinger operators. This allowed to describe explicitly the ground 
states of the corresponding operators (see \cite{CFKS},\cite{N83} for the details). 

In \cite{NV97} Novikov and one of the authors found a class of operators
with magnetic fields which are integrable on two different energy levels including the ground state.
In section 5 of this paper we present some generalisations of this result, but our main
goal here is to investigate what the factorisation method can give for the theory of 
Schr\"odinger operators $L$ with magnetic 
fields on a curved two-dimensional surface $M^2$. 

Although some ideas have been developed already in the 19-th century
by Darboux, Moutard et al (see e.g. \cite{D}), it was probably Schr\"odinger who first used the 
factorisation method in quantum mechanics \cite{Sch} (see also \cite{IH}). The idea of this method 
is very simple:
if a given operator $L$ can be factorised as
$$
{ L} = {D}_1 { D}_2
$$
then the new operator
$$
\tilde { L} = { D}_2 {D}_1
$$
has the same spectrum as ${ L}$ (provided the operators are good enough) except possibly $\lambda =0$ 
and if one knows the eigenfunctions $\psi$ of $L$ then the formula
$$
\tilde \psi=D_2 \psi
$$
gives the eigenfunctions of $\tilde { L}$. One can obviously include a shift into the factorisation scheme:
$$
L=D_1D_2+c \to \tilde L=D_2D_1+c
$$
where $c$ is a constant. For one-dimensional Schr\"odinger operator it is always possible to 
continue this procedure infinitely and construct a chain of operators related by this transformation
usually called the {\it factorisation chain} (or dressing chain). In the case when this chain is periodic 
the spectrum and eigenfunctions of all these operators can be described explicitly (see \cite{SV}).
The classical example is harmonic oscillator when the corresponding $\tilde L=L+const$.

In two dimensions the factorisation of the general Schr\"odinger operator $L$ with magnetic field
on a curved surface $M^2$
is possible only in the special case when the potential is equal up to a sign to the magnetic field
(see Theorem 1 below). A simple but important calculation shows that the transformation
$$
L=D^*D \to \tilde L=DD^*
$$
changes the magnetic field $B$ by the Gaussian curvature $K$ of $M^2$:
$$
\tilde B=B+K.
$$
In contrast to the one-dimensional case in two dimensions one can not in general continue the 
factorisation procedure. We give a complete classification of all possible factorisation 
chains on a curved surface $M^2$ (Theorem 2) which in particular says that infinite factorisation chain 
exists only on the surfaces with constant Gaussian curvature. The magnetic field and the potential 
of the corresponding operators must also be constant. 

In section 3 we discuss the global geometric aspects of the factorisation chains.
The Index theorem and the classical Gauss-Bonnet formula play here a crucial role. 
They explain a big difference between positive and negative curvature cases.
In the positive constant curvature case when $M^2=S^2$ is the standard round sphere we have the 
{\it Dirac magnetic monopole} \cite{Dir31}, \cite{Tamm}, \cite{WY}. 
We show how factorisation method leads 
to the complete description of the 
spectrum of the corresponding Schr\"odinger operator $L$ in the same way as
Schr\"odinger did for the harmonic oscillator in \cite{Sch}.
For the flat torus we have the standard {\it Landau problem} \cite{LL}.
We also discuss what the factorisation method gives for the analogue of the Landau problem 
on a surface of constant negative curvature with genus more than 1.

In section 4 we present some new examples of integrable Schr\"odinger operators 
with magnetic fields related to the two-term factorisation chains on the surfaces 
with non-constant curvature. 
The main observation here is that if the 
Laplace-Beltrami operator on $M^2$ is integrable then the same is true for the operator with the additional 
magnetic field $B=\pm K$ and the potential $U=K$, $K$ being the Gaussian curvature. 

In the last section we consider the quasi-cyclic chains of the Schr\"odinger
operators with magnetic fields on a curved surface related by Laplace transformations
generalising the constructions from the paper \cite{NV97}. This leads to a class of
the operators with two known energy levels, one of which is the ground state.

\section{Factorisation method on curved surfaces: local theory}

Consider an oriented analytic surface $M^2$ with a Riemannian metric $ds^2$. It is known that $M^2$ has a complex structure 
such that the metric is conformal. In any complex chart $z=x+iy, \ \bar z=x-iy$ the metric has the form
$$
ds^2=\frac{dz d\bar z}{h^2(z, \bar z)}=\frac{dx^2+dy^2}{h^2(z, \bar z)}.
$$
The Laplace-Beltrami operator $\Delta_h$ can be defined locally as
$$
\Delta_h= 4h^2 \ \partial\bar \partial=h^2 \ ({\partial _x^2}+{\partial _y^2})
$$
where $\partial={\partial _z}=\frac{1}{2}({\partial _x}-
i{\partial _y}), \ 
\bar \partial=\partial _{\bar z}=\frac{1}{2}({\partial _x}+
i{\partial _y})$. To introduce magnetic field one should replace the usual derivatives by their covariant counterparts:
\begin{equation}
\begin{array}{c}
\nabla_x=\partial _x-ia, ~~~ \nabla_y=\partial _y-ib, \\
\ \\
\nabla=\partial -iA, ~~~ \bar \nabla=\bar \partial-i\bar A, \\
\ \\
A=\frac{1}{2}(a-ib), ~~~ \bar A=\frac{1}{2}(a+ib).
\end{array}
\label{3}
\end{equation}
The corresponding Schr\"odinger operator
$$
L_A=-h^2\ [(\partial_x-ia)^2+(\partial_y-ib)^2]
$$
can be rewritten as
\begin{equation}
L_A=-4\ h^2\ \nabla\bar \nabla+h^2H=-4\ h^2\ \bar \nabla \nabla-h^2H=-2\ h^2\ (\nabla \bar \nabla+\bar \nabla \nabla)
\label{5}
\end{equation}
where
$$
H=\partial_xb-\partial_ya=i\ [\nabla_x, \nabla y]=2\ [\nabla, \bar \nabla].
$$
Geometrically, we have a complex $U(1)$-bundle over $M^2$ with the connection form
\begin{equation}
\alpha=i\ (adx+bdy)
\label{6}
\end{equation}
and the curvature
\begin{equation}
\Omega=d\alpha=i\ H \ dx\wedge dy=-\frac{H}{2} \ dz\wedge d\bar z
\label{7}
\end{equation}
(see the next section for further discussion of the geometric aspects).
Let's define magnetic field $B$ by the relation
\begin{equation}
\Omega=iB \ d\sigma
\label{8}
\end{equation}
where $d\sigma=\frac{1}{h^2} \ dx\wedge dy$ is the area element of the surface.
By definition we have
\begin{equation}
B=h^2H
\label{9}
\end{equation}
The most general Schr\"odinger operator on $M^2$ has the form
\begin{equation}
L=-h^2\ [(\partial_x-ia)^2+(\partial_y-ib)^2]+U,
\label{10}
\end{equation}
where the potential $U(x, y)$ is a real function on $M^2$.

Consider the following {\it factorisation problem} for (\ref{10}): when $L$ can be represented locally
as
\begin{equation}
L=(\alpha_1\partial+\alpha_0)(\alpha_1^{*}\bar \partial+\alpha_0^{*})
\label{11}
\end{equation}
($\alpha$-factorisation) or
\begin{equation}
L=(\beta_1\bar \partial+\beta_0)(\beta_1^{*} \partial+\beta_0^{*})
\label{12}
\end{equation}
($\beta$-factorisation) for some functions $\alpha, \alpha^{*}, \beta, \beta^{*}$ ? 
In the Euclidean case such factorisations appeared in the theory of Pauli operators 
for spin 1/2 particles (see \cite{AC}, \cite{CFKS}). On a curved surface the situation
is pretty similar.

\begin{theorem}
The $\alpha$-factorisation for the Schr\"odinger operator (\ref{10}) exists iff $U=-B$. Similarly, the 
necessary and sufficient condition for the $\beta$-factorisation is $U=B$.
\end{theorem}
The sufficiency readily follows from formulae (\ref{5}):

\noindent if $U=-B$ then $L=-4\ h^2\  \nabla \bar \nabla$;

\noindent if $U=B$ then $L=-4\ h^2\  \bar \nabla  \nabla$.

The factorisation is not unique: if $L=D_1D_2$ is any factorisation, then
$L=(D_1f^{-1})(f D_2)$ is another one for an arbitrary function $f$. 
This actually gives all such factorisations.

Having a factorised operator $L=D_1D_2$ one can consider the new operator
$$
\tilde L=D_2D_1.
$$
Notice that the change of factorisation $D_1\to D_1f^{-1}, \ D_2\to f D_2$ corresponds to the gauge 
transformation of $\tilde L$:
$$
\tilde L\to fD_2D_1f^{-1}=f\tilde L f^{-1}.
$$
The magnetic field and the potential do not depend on the gauge and are defined correctly.
Let's compute them for
$\alpha$-factorised $L$. We can assume that $D_2=\bar \nabla, \ D_1=-4\ h^2\ \nabla$ so that
\begin{equation}
\begin{array}{c}
\tilde L=D_2D_1=-4\ \bar \nabla \ (h^2\nabla)=-4\ h^2\ \bar \nabla \nabla-2\ h_{\bar z}h\ \nabla= \\
\ \\
-4\ h^2\ (\bar \nabla +2h_{\bar z}h^{-1})\ \nabla=-4\ h^2\ \tilde{\bar \nabla}\tilde \nabla,
\end{array}
\label{16}
\end{equation}
where $\tilde{\bar \nabla}= \bar \nabla +2\ h_{\bar z}h^{-1}, \ \tilde \nabla=\nabla$.

{\bf Remark.} The nonsymmetry between $\tilde \nabla$ and $\tilde{\bar \nabla}$ can be easily corrected by a 
suitable gauge transformation:
$\tilde \nabla  \to h \tilde \nabla h^{-1} = \nabla - h_{z} h^{-1}, 
\tilde{\bar \nabla} \to h \tilde{\bar \nabla} h^{-1} = \bar \nabla + h_{\bar z}h^{-1}.$
In the future we will not worry about such a nonsymmetry provided the corresponding magnetic field is 
real.

The new magnetic field is 
$$
\begin{array}{c}
\tilde B=2\ h^2\ [\tilde \nabla, \tilde{\bar \nabla}]=2\ h^2\ [\nabla, \bar \nabla+2\ h_{\bar z}h^{-1}]=\\
\ \\
B+4\ h^2\ (h_{\bar z}h^{-1})_z=B+4\ h^2\ (\ln h) _{z\bar z}=B+h^2\triangle \ln h=B+K,
\end{array}
$$
where $K$ is the {\it Gaussian curvature} of the surface. Here we have used the standard formula for $K$ in 
the conformal coordinates
\begin{equation}
K=h^2\triangle \ln h,
\label{18}
\end{equation}
(see e.g. \cite{DNF}). Thus the new magnetic field is
\begin{equation}
\tilde B=B+K.
\label{15}
\end{equation}
According to the Theorem 1 the new potential is
\begin{equation}
\tilde U= \tilde B=B+K.
\label{15}
\end{equation}
Similarly, for $\beta$-factorisation the new magnetic field and the new potential are
$$
\tilde B=B-K, ~~~ \tilde U=K-B.
$$
Let's introduce the notation $(B, U)$ for the gauge class of Schr\"odinger operators (\ref{10})
with given magnetic field $B$ and potential $U$. Then we have the following two-term factorisation chains:
\begin{equation}
(B, -B)\stackrel{\alpha}{\to} (B+K, B+K)
\label{21}
\end{equation}
and
\begin{equation}
(B, B)\stackrel{\beta}{\to} (B-K, K-B).
\label{22}
\end{equation}
Notice that $\alpha \circ \beta=\beta \circ \alpha=Id:$
$$
\begin{array}{c}
(B, -B) \stackrel{\alpha}{\to} (B+K, B+K) \stackrel{\beta}{\to} (B, -B), \\
\ \\
(B, B) \stackrel{\beta}{\to} (B-K, K-B) \stackrel{\alpha}{\to} (B, B).
\end{array}
$$
The question is: can we continue the $\alpha$-chain allowing a shift by a constant? The answer is simple: 
in order to have $\alpha$-factorisation for an operator from the class $(B+K, B+K-2c), \ c=const$ we should require
$B+K-2c=-(B+K)$ or  $B+K=c$. In this case we have the following 3-term chains with an arbitrary constant $c$:
\begin{equation}
\begin{array}{c}
(c-K, K-c) \stackrel{\alpha}{\to} (c, c)=(c, -c)+2c \stackrel{\alpha}{\to} \\
\ \\
\stackrel{\alpha}{\to}(c+K, c+K)+2c=(c+K, 3c+K)
\end{array}
\label{23}
\end{equation}
and similarly
\begin{equation}
\begin{array}{c}
(c+K, c+K) \stackrel{\beta}{\to} (c, -c)=(c, c)-2c \stackrel{\beta}{\to} \\
\ \\
\stackrel{\beta}{\to}(c-K, K-c)-2c=(c-K, K-3c).
\end{array}
\label{24}
\end{equation}
Notice that the operators with constant magnetic field and constant potential 
can be considered as the natural analogues of the {\it Landau operators} on a curved surface $M^2$.
One more step in the factorisation procedure is possible only if the Gaussian curvature is constant:
$K=K_0$. In that case we actually can perform infinitely many steps
\begin{equation}
\begin{array}{c}
(c-K_0, \ K_0-c)\stackrel{\alpha}{\to} (c, c) \stackrel{\alpha}{\to} (c+K_0, \ 3c+K_0)
\stackrel{\alpha}{\to} ... \\
\ \\
\stackrel{\alpha}{\to}(c+mK_0, \ (2m+1)c+m^2K_0) \stackrel{\alpha}{\to} ...
\end{array}
\label{25}
\end{equation}
for any $m\in \bf{Z} _{+}$ and similarly
\begin{equation}
\begin{array}{c}
(c+K_0, \ c+K_0)\stackrel{\beta}{\to} (c, -c) \stackrel{\beta}{\to} (c-K_0, \ K_0-3c)
\stackrel{\beta}{\to} ... \\
\ \\
\stackrel{\beta}{\to}(c-mK_0, \ -(2m+1)c+m^2K_0) \stackrel{\beta}{\to} ....
\end{array}
\label{26}
\end{equation}
Summarising all this we have

\begin{theorem}
All possible factorisation chains on a surface $M^2$ with Gaussian curvature $K$ are given by

\noindent 1) two-term chains (\ref{21}), (\ref{22}) with an arbitrary magnetic field $B$;

\noindent 2) three-term chains  (\ref{23}), (\ref{24}) of the operators on $M^2$
with constant magnetic fields.

\noindent If a surface $M^2$ has a constant Gaussian curvature then we have also 

\noindent 3) infinite chains  (\ref{25}), (\ref{26}) with 
constant magnetic fields.
\end{theorem}

\section{Global geometry and spectral properties of $L$}

Let's assume now that $M^2$ is a closed surface of genus $g$
with given Riemannian metric, $\xi$ is a complex 
$U(1)$-bundle over $M^2$,
$\alpha$ is a connection on $\xi$,
$\Omega=iB \ d\sigma$ is its curvature form.
Here as above $d\sigma$ is the area element of the surface
determined by the metric and $B$ is a function on $M^2$ called magnetic field.

{\bf Remark.} We should mention that for a given magnetic field $B$ the corresponding connection 
$\alpha$ is defined uniquely modulo natural gauge transformations only when $M^2$
is a topological sphere. For a surface of genus $g$ one can always add to $\alpha$ 
a closed 1-form without changing $B$. Modulo exact forms corresponding to the gauge transformations
these forms form the first cohomology group $H^{1}(M^2,\bf{R}) \simeq {\bf{R}}^{2g}$.
These additional $2g$ parameters are called the {\it Aharonov-Bohm} fluxes
(see e.g. \cite{ASZ}).  All the spectra we consider in general depend not only on the 
magnetic field $B$ but also on the choice
of the connection $\alpha$ and therefore on these parameters.
When $g=1$ (i.e. when $M^2$ is a torus) this corresponds to 
the choice of $2$ {\it Bloch quasi-momenta}.

Notice that the total magnetic flux is an integer multiple of $2\pi$:
\begin{equation}
\frac{1}{2\pi}\int_{M^2}B\ d\sigma=b, ~~~ b\in {\bf{Z}}.
\label{27}
\end{equation}
The integer $b$ is actually the first Chern class of $\xi$:
\begin{equation}
b=c_1(\xi)\in H^2(M^2,\bf{Z}) \simeq {\bf{Z}}
\label{28}
\end{equation}
(see e.g. \cite{DNF}).
Having all this plus a potential $U$ which is a function on $M^2$ 
one can define the Schr\"odinger operator $L$ (\ref{10})
as explained in the previous section.
This operator is acting on the sections $S(\xi)$ of the bundle $\xi$
and is self-adjoint with respect to the natural Hermitian structure with the norm
$$
||\psi||^2 = \int_{M^2} |\psi|^2 \ d\sigma.
$$

Let's analyse now what happens when we apply the factorisation procedure.
Consider first the two-term chain (\ref{21}):
$$
(B, -B)\stackrel{\alpha}{\to} (B+K, B+K)
$$
or in the local coordinates
$$
L=D^{*}D\to DD^{*}=\tilde L
$$
where $D=2\bar \nabla, \ D^{*}=-2h^2\nabla$.
Here we have used the complex structure on $M^2$ uniquely determined by the Riemannian metric, 
which exists according to the classical results (see e.g.\cite{DNF}). 

Notice first of all that the operator $D$ maps the 
sections $S(\xi)$ of $\xi$ into the sections $S(\eta)$ of the bundle $\eta = \xi\otimes T^{0,1}(M^2)$
which are the antiholomorphic 1-forms on $M^2$ with the values in $\xi$ of the form $\psi d \bar z, 
\psi \in S(\xi)$. This space also has the natural Hermitian structure induced by the Hermitian structure on 
$\xi$
and the metric on $M^2$. It is easy to check that $D^{*}=-2h^2\nabla$ is indeed operator adjoint to 
$D=2\bar \nabla$
with respect to these structures, so the operator $L=D^{*}D$ is non-negative and
$Ker L = Ker D$.

The new operator $\tilde L = D D^{*}$ is acting on the sections $S(\eta)$ of the bundle $\eta$ and 
in the local coordinates has the form (\ref{16}):
$$\tilde L=-4\ \bar \nabla \ (h^2\nabla)= -4\ h^2\ (\bar \nabla +2h_{\bar z}h^{-1})\ \nabla.$$
The new covariant derivatives $\tilde{\bar \nabla}= \bar \nabla +2\ h_{\bar z}h^{-1},
\ \tilde \nabla=\nabla$ correspond to the natural connection on $\eta$ induced by the connection
$\alpha$ on $\xi$ and the natural Hermitian connection on the antiholomorphic cotangent bundle
$T^{0,1}(M^2)$ (see \cite{GH}). The curvature of this connection is
$$\tilde \Omega = i(B+K)d\sigma,$$
where $K$ is the Gaussian curvature of the metric.
Notice that the total flux of the new magnetic field $\tilde B = B + K$
(or, equivalently, the first Chern class of the bundle $\eta$) is
\begin{equation}
c_1(\eta) = \tilde b=\frac{1}{2\pi}\int_{M^2}\tilde B\ d\sigma=b+\chi=b+(2-2g)
\label{30}
\end{equation}
where $\chi=2-2g$ is the {\it Euler characteristics} of $M^2$ because of the {\it Gauss-Bonnet formula}
\begin{equation}
\int_{M^2} K\ d\sigma=2\pi \chi.
\label{31}
\end{equation}

The operator $D$ is an elliptic operator from  $S(\xi)$ into $S(\eta)$. Its 
index can be determined by the Index theorem and is given by the Riemann-Roch formula
\begin{equation}
{\rm ind} \ D={\rm dim \ Ker \ } D-{\rm dim \  Ker \ } D^{*}=b-g+1.
\label{32}
\end{equation}
When  magnetic field is large enough, more precisely when $b>2g-2$, we have 
\begin{equation}
{\rm dim \  Ker \ } D^{*}=0, ~~~ {\rm dim \ Ker \ } D=b-g+1.
\label{36}
\end{equation}
In that case the ground state of the operator $L$ is degenerate:
$$
{\rm dim \  Ker \ } L={\rm dim \  Ker \ } D=b-g+1
$$
while $\tilde L$ is positive operator:
$$
{\rm dim \  Ker \ } \tilde L={\rm dim \  Ker \ } D^{*}=0.
$$
The rest of the spectrum (which is discrete according to  general theory) is the same for $L$ and
$\tilde L$: the intertwining operators $D$ and $D^{*}$ establish the isomorphism of the 
corresponding eigenspaces. Notice that if $b=g-1$ then, according to the Riemann-Roch formula (\ref{32}),
${\rm dim \ Ker \ } L={\rm dim \ Ker \ } \tilde L$ so that $L$ and $\tilde L$ are isospectral everywhere
( not just for $\lambda>0$). Now for the general factorisation chain one should only take into account
the additional shift of the spectrum.

For the $\beta$-factorisation chains  (\ref{22}) the analysis is similar, one should simply 
replace the holomorphic structure by the antiholomorphic one.

Let's consider now some examples.

\bigskip

{\bf Example 1. Dirac magnetic monopole on a sphere.}

Let $M^2$ be a sphere $S^2\subset R^3$ with the standard metric of constant Gaussian curvature $K$.  
The Hamiltonian of the Dirac monopole with a charge $q\in \bf{Z}$ in our notations is 
a Schr\"odinger operator 
$H_q$ from the gauge class 
$(B, 0)$ where $B$ is a constant satisfying the quantisation relation
$$
\frac{1}{2\pi}\int_{S^2} B \ d\sigma = \frac{B}{2\pi} \int_{S^2}d\sigma=2B=q,
$$
i.e. $B=q/2$ must be integer or half-integer. 
It is acting on the sections of the U(1)-bundle with the first Chern class $q$.
As we have seen above such operator can be included in the 
infinite factorisation chain
\begin{equation}
\begin{array}{c}
(B, 0)=(B, -B)+B\to (B+K, \  B+K)+B=(B+K, \  2B+K)= \\
\ \\
(B+K, \  -(B+K))+3B+2K \to (B+2K, \ B+2K)+3B+2K= \\
\ \\
(B+2K, \ 4B+4K)\to  \ ... \ 
\to (B+mK, \ 2mB+m^2K) \to ...,
\end{array}
\label{38}
\end{equation}
$m\in \bf{Z}_{+}$.
If $B=q/2>0$ then the operator $(B, -B)$ is positive and according to index theorem (\ref{36})
$$
{\rm dim \  Ker \ } (B, -B)=q-g+1=q+1
$$
since $g=0$ and $q\geq 2g-1=-1$. Therefore the ground state of $H_q$ has the energy $\lambda=B=q/2$ and 
the corresponding eigenspace has dimension 
$q+1=2B+1$. The second operator $(B+K, 2B+K)$  in the chain has the same spectrum as $(B, 0)$ except the ground state.
By the same reasons its ground state has energy $\lambda=3B+2K$ which is degenerate:
$$
\begin{array}{c}
{\rm dim \ Ker \ } [(B+K, \  2B+K)-(3B+2K)]={\rm dim \ Ker \ } [B+K,\  -(B+K)]= \\
\ \\
2B+2+1=2B+3.
\end{array}
$$
Thus the second eigenvalue of $H_q$ is $\lambda_2=3B+2K$ with degeneracy $2B+3$. 
On the m-th step we'll have the operator
$(B+mK, \ 2mB+m^2K)=(B+mK, \ -B-mK)+(2m+1)B+(m^2+m)K$ which leads to the eigenvalue 
$$
\lambda_m=(2m+1)B+m(m+1)K
$$
with degeneracy $2B+2m+1$. Thus we arrived at the well-known result about the spectrum of the Dirac monopole
(see e.g. \cite{WY}):
$$
{\rm Spec \ } H_q=\{\lambda_n=(2m+1)\frac{q}{2}+m(m+1)K ~~ {\rm with \ degeneracy} ~~
q+2m+1\},
$$
$ m=0, 1, 2,...$

We can make all this explicit (including the calculation of the corresponding eigenfunctions) 
using the stereographic coordinate $z=x+iy$. Assume for simplicity that the radius of the sphere $R=1$,  
so that the Gaussian curvature $K=1/R^2=1$. The metric has the form
$$
ds^2=\frac{4}{(1+z\bar z)^2} \ dz d\bar z.
$$
The corresponding chain of operators in a suitable gauge has the form
\begin{equation}
\begin{array}{c}
 L_N=-(1+z\bar z)^2 \ \partial \bar  \partial-Nz(1+z\bar z) \ \partial+N\bar z (1+z\bar z) \ \bar \partial
+N^2(1+z\bar z)= \\
\ \\
D_N^{*}D_N+N(N+1)=D_{N-1}D_{N-1}^{*}+N(N-1), \\
\ \\
D_N=(1+z\bar z) \ \bar \partial +Nz; ~~~ D_N^{*}=-(1+z\bar z) \ \partial +(N+1)\bar z.
\end{array}
\label{40}
\end{equation}
Here
$
L_N=H_{2N}+N^2,
$
where $H_{2N}$ is the Dirac monopole operator with the charge $q=2N$, $B=N$. We obviously have the 
intertwining relations
\begin{equation}
L_{N+1} D_N=D_N L_N, ~~~ L_N D_N^{*}=D_N^{*} L_{N+1}.
\label{41}
\end{equation}
To find the ground state of $L_N: \  L_N\psi=N(N+1)\psi$ one should solve the equation
$D_N\psi=0$:
$$
(1+z\bar z) \ \bar \partial \psi+Nz \ \psi=0 , ~~ {\rm or} ~~ \bar \partial \ \ln \psi=-\frac{Nz}{1+z\bar z}
$$
The solutions are easy to find:
$$
\psi=\frac{f(z)}{(1+z\bar z)^N}
$$
where $f$ is any holomorphic function of $z$. Because of the condition
$$
\int\vert\psi \vert^2 d\sigma=\int \int_{R^2} 
\frac{\vert\psi \vert^2 dzd\bar z}{(1+z\bar z)^2}<\infty
$$
the function $f(z)$ must be a polynomial of order $\leq 2N$. This gives us the space of dimension $2N+1$. 
Applying to this space the ``lowering" operators $D_{N-1}^{*}, D_{N-2}^{*}, ...$ we will construct all 
eigenfunctions of the operators $L_{N-1}, L_{N-2}, ...$ with the eigenvalue $\lambda = N(N+1)$. 
This gives the following description of the eigenfunctions of the Dirac monopole operator $H_q$.

\begin{theorem}

The eigenfunctions of the Dirac magnetic monopole operator with the charge 
$q$ on a unit sphere corresponding to the eigenvalue
\begin{equation}
\lambda=(2m+1)\frac{q}{2}+m(m+1), ~~~ m\in Z_{+}
\label{43}
\end{equation}
form the space of dimension $q+2m+1$ which can be described as 
\begin{equation}
\psi=D_{N-m}^{*} \  ... \  D^{*}_{N-2} D_{N-1}^{*} \ \frac{f(z)}{(1+z\bar z)^N}
\label{44}
\end{equation}
where $N=m+\frac{q}{2}$ and $f(z)$ is an arbitrary polynomial of degree
${\rm deg} f\leq 2N=2m+q$. 

\end{theorem}

One can check that the formulas for $\psi$ we have given actually determine the smooth sections of
the corresponding line bundles over the sphere $S^2$.

The eigenfunctions of the Dirac magnetic monopole are known as {\it monopole harmonics} 
and have been investigated by Wu and Yang \cite{WY}, who were probably the first to identify them 
explicitly as the sections. Our derivation is different and more close to the one of the paper 
\cite{DV} by D'Hoker and Vinet who discovered the supersymmetry of the corresponding Pauli 
equation in the presence of a magnetic monopole (see also \cite{JMPH}).

{\bf Remark.} If $q=2N$ is  even integer then $H_{2N}=L_N-N^2$ can be intertwined with the shifted
standard Laplace-Beltrami operator $L_0 = -\Delta_g$ on the sphere $S^2$:
$$
H_{2N} D=D (L_0-N^2), ~~~ D=D_{N-1} \ ... \ D_1 D_0,
$$
so that one can use the well-known eigenfunctions of $L_0$ (spherical harmonics) to construct the 
eigenfunctions of $H_q$. Notice that we have given an alternative description of spherical harmonics using 
the factorisation chain. The explicit form of the intertwining operator $D$ is 
$$
D=\frac{1}{(1+z\bar z)^{N-2}} \ \bar \partial \ (1+z\bar z)^2 \ \bar \partial \ (1+z\bar z)^2 \ 
\bar \partial \ ...
\ \bar \partial \ (1+z\bar z)^2 \ \bar \partial
$$
which is of the order $N$ in $\bar \partial$.

It is interesting to compare the Dirac monopole problem on a sphere with the Landau problem on the
surfaces of genus $g\geq 2$ with the constant negative Gaussian curvature $K<0$
which we analyse below. The corresponding classical problems behave very differently 
so one should expect the same for the quantum problems as well.
 
For the investigation of the intermediate case of flat torus which is the classical Landau problem
we refer to \cite{LL} and \cite{N83}. In that case $K=0$ and our formula (\ref{38}) leads to the 
usual Landau spectrum
$$
\lambda_m=\{(2m+1)B, ~~~ m=0, 1, ...\},
$$
where $B$ is assumed positive and quantised.
The corresponding eigenfunctions can be expressed in terms of the classical elliptic $\sigma$-functions
(see \cite{N83}, \cite{DN80} for the details).

\bigskip

{\bf Example 2. Landau problem on a surface of genus $g\geq 2$.}

Let $M^2$ be any analytic surface with the metric $ds^2$ of constant curvature $K=-1$. Such surface $M^2$ 
can be considered as a quotient
of the Lobachevsky plane $\cal L$ by an infinite discrete group $G$:
$$
M^2={\cal L}/G.
$$
${\cal L}$ can be realised as an open disc $\vert z \vert < 1$ with the metric
$$
ds^2=\frac{4}{(1-z\bar z)^2} \ dz d\bar z.
$$
The Schr\"odinger operator on ${\cal L}$  with constant magnetic field $B$ (Landau operator) 
can be written as
$L_B^{\cal L}=-(1-z\bar z)^2 \ (\nabla \bar \nabla+ \bar \nabla \nabla)$, where
$$
\nabla=\partial-B\frac{\bar z}{1-z\bar z}, ~~~   \bar  \nabla=\bar \partial+B\frac{ z}{1-z\bar z},
$$
or, explicitly,
\begin{equation}
L_B^{\cal L}=-(1-z \bar z)^2\ \partial \bar \partial  +B\bar z(1-z \bar z) \ \bar \partial
-Bz(1-z \bar z) \ \partial+B^2z \bar z.
\label{45}
\end{equation}
In order to define the corresponding operator $L_B$ acting on the sections of some 
$U(1)$-bundle over $M^2$ one needs the
quantisation condition 
\begin{equation}
\frac{1}{2\pi}\int_{M^2}B \ d\sigma =\frac{1}{2\pi} B\int_{M^2} \ d\sigma=(2g-2)B \in \bf{Z}
\label{46}
\end{equation}
to be satisfied.
Here we have again used the Gauss-Bonnet formula  
$$
\frac{1}{2\pi}\int_{M^2} K\ d\sigma=-\frac{1}{2\pi}\int_{M^2} \ d\sigma=\chi=2-2g.
$$
Notice that to define $L_B$ one also needs to choose a connection $\alpha$ which depends on
$2g$ Aharonov-Bohm fluxes (see the remark at the beginning of this section).

Let's see what the factorisation chain (\ref{25}) gives us for the calculation of the spectrum of $L_B$.
We have the chain (\ref{38}) with $K=-1$:
$$
\begin{array}{c}
(B, 0)\to (B, -B)+B\to (B-1, B-1)+B=(B-1, 2B-1)\to ... \\
\ \\
\to (B-m, 2mB-m^2)=(B-m, m-B)+(2m+1)B-(m^2+m).
\end{array}
$$
Let's assume that $B>0$. By Riemann-Roch formula (\ref{32}) the index of the operator
$(B-m, m-B)$ is $(2g-2)(B-m)-g+1=(2g-2)(B-m-\frac{1}{2})$, so it is positive if
$$
m<B-\frac{1}{2}.
$$
Using the same arguments as in the previous example we can claim that the first $[B-\frac{1}{2}]+1$
eigenvalues of the operator $L_B$ have the form
\begin{equation}
\lambda_m=(2m+1)B-m(m+1), ~~~ m=0, 1, 2, ..., [B-\frac{1}{2}].
\label{48}
\end{equation}
Let's remind that $B$ has the form ${k}/(2g-2)$ for some positive integer $k$. Moreover, we can say that 
if $(2g-2)(B-m)>2g-2$, i.e. if
$$
m<B-1
$$
then the corresponding eigenspace has the dimension
\begin{equation}
{\rm dim \ Ker \ } (L_B-\lambda_m)=(2g-2)(B-m-\frac{1}{2}).
\label{50}
\end{equation}
Notice that the spectrum
(\ref{48}) depends only on the magnetic field $B$ but not on the Aharonov-Bohm fluxes.
It is related to the discrete part of the spectrum of the Landau operator $L_B^{\cal L}$
on the whole Lobachevsky plane (see e.g.\cite{W}). It would be interesting to find an effective 
representation for the corresponding eigenfunctions
on a surface $M^2$ given explicitly as an algebraic curve in $\bf{C}^2$.

From these considerations we have nothing to say about the rest of the spectrum of $L_B$
(which in fact depends both on Aharonov-Bohm fluxes and on $(3g-3)$ complex parameters
({\it moduli}) determining the conformal structure on $M^2$). In
particular, we can not say anything about the spectrum of the pure Laplace-Beltrami operator 
$L_0$ on $M^2$. This is a reflection of the fact that the corresponding classical geodesics 
problem on $M^2$ is non-integrable. 

\bigskip

\section{New examples of the integrable quantum problems with magnetic field}

Let's consider any surface $M^2$ with integrable quantum geodesic problem
$$
L \psi = \lambda \psi,
$$
where $L = -\Delta_h$ and $\Delta_h$ is the Laplace-Beltrami operator on $M^2$.
By definition this means that there exists another differential operator $F$ which commutes with $L$: $[F, L]=0$
and has an independent highest symbol. We have shown (see (\ref{23}) above) that $L$ can be factorised $L=D^{*}D$ and the 
new operator$\tilde L=DD^{*}$ has the magnetic field $B= \pm K$ ($K$ is the Gaussian curvature) 
and the potential $U=K$. We claim that this new operator is also integrable. 
Indeed, consider the differential operator
\begin{equation}
\tilde F=DFD^{*},
\label{51}
\end{equation}
then
$$
\tilde L \tilde F=DD^{*}DFD^{*}=DLFD^{*}=DFLD^{*}=DFD^{*}DD^{*}=\tilde F \tilde L,
$$
i.e. $\tilde F$ commutes with $\tilde L$. It is easy to check that if the highest symbol of $F$ is independent of the highest symbol
of $L$ then the same is true for $\tilde F$ and $\tilde L$. The intertwining relations
$$
\tilde LD=DL ~~~ {\rm and} ~~~ D^{*}\tilde L=LD^{*}
$$
establish isomorphism between the two spectral problems
$$
L\psi=\lambda \psi ~~~ {\rm and} ~~~ \tilde L \tilde \psi=\lambda \tilde \psi
$$
for any $\lambda\ne 0: \ \tilde \psi=D\psi, \ \psi = D^{*} \tilde \psi$. For $\lambda=0$ the situation is 
described by the Riemann-Roch formula:
\begin{equation}
{\rm dim \ Ker \ } L-{\rm dim \ Ker} \  \tilde L = 1-g
\label{52}
\end{equation}
In particular, for the sphere $S^2$ we have $g=0$,
$$
{\rm dim \  Ker \ } \ L=1, ~~~ {\rm dim \ Ker \ } \ \tilde L=0.
$$
This means that the ground state of $\tilde L$ has the energy which is equal to the minimal positive eigenvalue of
$L$: if ${\rm Spec} \  L=\{\lambda_0=0, \lambda_1, \lambda_2, ... \}$, then  
${\rm Spec} \ \tilde  L=\{\lambda_1, \lambda_2, ... \}$. 

For the torus $T^2$ we have $g=1$ and 
$$
{\rm dim \ Ker \ } L= {\rm dim \ Ker \ } \tilde L=1
$$
so that
$$
{\rm Spec \ L}= {\rm Spec \ } \tilde L.
$$
This is true for all the values of quasi-momenta i.e. for the whole Bloch spectrum.
Notice for the torus the total magnetic flux of $\tilde L$ is zero.

We do not know any integrable Laplace-Beltrami operators on a surface of genus $g\geq 2$. 
For the corresponding classical problem about geodesics on $M^2$ there exists a rigorous proof 
that there are no such metrics
(see \cite{Kozlov}).

\begin{theorem}
Let $M^2$ be any surface such that the corresponding quantum geodesic problem
$-\Delta_h \psi = \lambda \psi$  is integrable. Then the Schr\"odinger operator $\tilde L$ 
on $M^2$ with magnetic field $B= \pm K$ and potential $U=K, \ K$
is the Gaussian curvature of $M^2$, is integrable too and has the same spectrum as $L=-\Delta_h$
with the only possible exception at $\lambda = 0$.
\end{theorem}

{\bf Example.} Consider an ellipsoid $M^2$
$$
\frac{x^2}{a^2}+\frac{y^2}{b^2}+\frac{z^2}{c^2}=1
$$
with the metric induced from $\bf{R}^3$. The geodesic problem on $M^2$ has been solved by Jacobi who showed
that it can be integrated by separation of variables. The same is true for the corresponding quantum problem.
Gaussian curvature of $M^2$ has the form
$$
K=(abc)^{-2}\left(\frac{x^2}{a^4}+\frac{y^2}{b^4}+\frac{z^2}{c^4}\right)^{-2}
$$
so we can claim that the Schr\"odinger operator $\tilde L$ on $M^2$ with the magnetic field
$$
B=(abc)^{-2}\left(\frac{x^2}{a^4}+\frac{y^2}{b^4}+\frac{z^2}{c^4}\right)^{-2}
$$
and the potential
$$
U=-(abc)^{-2}\left(\frac{x^2}{a^4}+\frac{y^2}{b^4}+\frac{z^2}{c^4}\right)^{-2}
$$
is integrable and ${\rm Spec} \ L={\rm Spec} \ \tilde L \cup \{ 0 \}$. Notice that the order of the additional quantum integral 
$\tilde F=DFD^{*}$ is 4, since the order of $F$ is known to be 2. 

{\bf Remark.} We do not claim that that the minimal order of the additional integral
is 4. On the contrary, in this case one can show that there exists an additional integral
of order 2. It would be interesting to investigate the corresponding classical mechanical problem
of motion on the ellipsoid in this special magnetic field. We conjecture that it is nonintegrable.

This example can be generalised in the following way.
Consider any surface $M^2$ with the Liouville metric
$$
ds^2=g_{11}du^2+g_{22}dv^2, ~~~~~ g_{11}=\frac{u-v}{f}, \ \ g_{22}=\frac{v-u}{g},
$$
where $f(u)$ and $g(v)$ are arbitrary functions of the specified arguments. 
Its Gaussian curvature is expressed by the formula
$$
K=\frac{f-g}{2(u-v)^3}-\frac{f'+g'}{4(u-v)^2}.
$$
The corresponding Laplace-Beltrami operator 
$$
\Delta_h = \sqrt {g^{11}g^{22}}\ \partial_u \ \frac{g^{11}}{\sqrt{g^{11}g^{22}}}
\ \partial_u +
\sqrt {g^{11}g^{22}}\ \partial_v \ \frac{g^{22}}{\sqrt{g^{11}g^{22}}}
\ \partial_v
$$
commutes with the second-order operator
$$
F= v \sqrt {g^{11}g^{22}}\ \partial_u \ \frac{g^{11}}{\sqrt{g^{11}g^{22}}}
\ \partial_u +
u \sqrt {g^{11}g^{22}}\ \partial_v \ \frac{g^{22}}{\sqrt{g^{11}g^{22}}}
\ \partial_v,
$$
so that variables can be separated both in the classical and quantum cases. 
So we can claim that the Schr\"odinger operator with magnetic field $B= \pm K$
and the potential $U=K$ is integrable on any Liouville surface.

As a degenerate case of this construction one can get the surfaces of revolution 
in $R^3$ with the metric
$$
ds^2=\rho(z)^2\ d\varphi^2+(1+\rho'(z)^2) \ dz^2.
$$
In that case the Gaussian curvature is given by the formula
$$
K=\frac{\rho' \rho''}{\rho(1+(\rho')^2)^2}.
$$

We would like to mention that there exist the intertwining relations between two Schr\"odinger operators
which are not related to any factorisation of the operators (see \cite{F} for some examples).
We will present here another example of this type modifying the previous analysis of the
Dirac monopole.
Recall that the operator $L_N$ of  Dirac monopole on the unit sphere  given by  (\ref{40})
$$
 L_N=-(1+z\bar z)^2 \ \partial \bar \partial-Nz(1+z\bar z) \ \partial+N\bar z (1+z\bar z) \
\bar \partial
+N^2(1+z\bar z)
$$
 satisfies the interwining relations (\ref{41})
$$
L_{N+1} D_N=D_N L_N
$$
where
$$
D_N=(1+z\bar z) \ \bar \partial +Nz.
$$
It can be readily verified that the modified operators $\tilde L_N$ given by
$$
\tilde L_N= L_N-\frac{1}{4}\ P^2-(N+1) \ P, ~~~ \tilde L_{N+1}=L_{N+1}-\frac{1}{4}\ P^2-N\ P
$$
satisfy the same intertwining relations 
$$
\tilde L_{N+1} \tilde D_N=\tilde D_N \tilde L_N
$$
where the modified intertwining operator $\tilde D_N$ is of the form
$$
\tilde D_N=D_N+\frac{\varphi}{1+z\bar z}.
$$
Here $\varphi$ is a quadratic polynomial in $z$: 
$$
 \varphi=q+pz-\bar qz^2, ~~~  p\in R, \  \ q, \bar q \in C,
$$
and $P$ is given by the formula
$$
P=\varphi'-\frac{2\bar z \varphi}{1+z\bar z}=p\ \frac{1-z\bar z}{1+z\bar z}
-2\ \frac{\bar q z+q \bar z}{1+z\bar z}.
$$
Geometrically, $P$ represents the restriction to the sphere $S^2$ 
of an arbitrary linear function from the ambient space
$R^3$. Without any loss of generality we may assume $q=\bar q=0$ (by appropriately choosing the axis
of  stereographic projection), so $P = p\ \frac{1-z\bar z}{1+z\bar z}$.

In particular for $N=1$ we have the intertwining relation between the Dirac monopole operator
with charge 2 in the potential $U = -\frac{1}{4}\ P^2 $
and the usual Laplace-Beltrami operator with the additional potential $V=-\frac{1}{4}\ P^2- \ P$.

One can show that this construction actually gives all the potential deformations of the intertwining relations
(\ref{41}).

\section{Laplace transformations on a curved surface and quasi-cyclic chains}

Let $L$ be any Schr\"odinger operator with magnetic field $B$ and potential $U$:
$$
L=-4 \ h^2 \ \nabla \bar \nabla+(U+B)=-4 \ h^2 \ \bar \nabla \nabla+(U-B).
$$
In general none of these forms is pure factorisable but on the level $\lambda =0: \ L\psi =0$ 
we still can do the transformation
$$
\tilde \psi =\nabla \psi ~~~ {\rm or} ~~~ \tilde \psi = \bar 
\nabla \psi.
$$
In particular, if 
$$
L\psi=[-4 \ h^2 \ \bar \nabla \nabla+(U-B)]\ \psi=0
$$
and $\tilde \psi = \nabla \psi$ we have
$$
\bar \nabla \tilde \psi = \frac{U-B}{4h^2} \ \psi
$$
and therefore
$$
\tilde \nabla \bar \nabla \tilde \psi = \frac{U-B}{4h^2} \ \tilde \psi
$$
where $\tilde \nabla$ satisfies the relation
$$
\tilde \nabla  \frac{U-B}{4h^2} = \frac{U-B}{4h^2} \ \nabla
$$
implying
\begin{equation}
\tilde \nabla = \nabla -\ln(U-B) _z+(\ln h^2)_z.
\label{100}
\end{equation}
Thus $\tilde \psi=\nabla \psi$ satisfies the new Schr\"odinger equation $\tilde L \tilde \psi =0,$
where
\begin{equation}
\tilde L=-4 \ h^2 \ \tilde \nabla \bar \nabla+(U-B)=-4 \ h^2 \  \tilde \nabla \bar \nabla+(\tilde U + \tilde B).
\label{101}
\end{equation}
The new magnetic field is
\begin{equation}
\tilde B=2\ h^2 \ [\tilde \nabla, \bar \nabla ]=B+2 \ h^2 \ (\ln (U-B))_{z\bar z}-
4 \ h^2 \ (\ln h)_{z \bar z}=
B+\frac{1}{2} \ \Delta_h(\ln (U-B)) -K
\label{102}
\end{equation}
where $\Delta_h=4h^2 \ \partial\bar \partial$ is the Laplace-Beltrami operator on $M^2$, 
$K$ is the Gaussian 
curvature. The new potential
is
\begin{equation}
\tilde U=U-B-\tilde B=U-[2B+\frac{1}{2} \ \Delta_h(\ln (U-B)) -K].
\label{103}
\end{equation}
The formulae (\ref{100},\ref{101},\ref{102},\ref{103}) define correctly the Laplace transformation
for the Schr\"odinger operators on a curved surface.
The only difference with the standard formulae in the flat case (see e.g. \cite{NV97}) 
is the additional Gaussian curvature term.

Following \cite{NV97} let's call the chain of Laplace transformations
\begin{equation}
\begin{array}{c}
B_{k+1}=B_k+\frac{1}{2} \ \Delta_h(\ln (U_k-B_k)) -K, \\
\ \\
U_{k+1}=U_k-B_k-B_{k+1}=U_k-[2B_k+\frac{1}{2} \ \Delta_h(\ln (U_k-B_k)) -K],
\end{array}
\label{*}
\end{equation}
$k=0, ..., N$, {\it quasi-cyclic} if both the initial and final Schr\"odinger operators are 
factorisable,
possibly at the different energy levels:
\begin{equation}
U_0+B_0=0, ~~~ U_{N}+B_{N}=U_{N-1}-B_{N-1}=-c
\label{**}
\end{equation}
where $c>0$ is a constant. 

As well as in the flat case \cite{NV97} the last operator $L_{N}$ has two ``integrable" levels:
$L_N\psi_0=0$ and $L_N \psi_c=-c \ \psi_c$. Indeed, $\psi_0$ can be found as the result 
$$
\psi_0 = \nabla_{N-1} \nabla_{N-2}...\nabla_{0} \psi
$$
of the Laplace transformations 
applied to the solutions of the initial equation $L_0\psi=0$, which is equivalent to 
\begin{equation}
\bar \nabla_0 \psi =0,
\label{A}
\end{equation}
while $\psi_c$ are the ground states of the operator  $L_N = -4 \ h^2 \ \nabla_N \bar \nabla_N - c$
satisfying the equation
\begin{equation}
\bar\nabla_N\psi=0, ~~~.
\label{B}
\end{equation}
Obviously we should assume that solutions of both (\ref{A}) and (\ref{B}) do exist, which imposes 
some global assumptions on the magnetic field.
Let's discuss these assumptions.

First of all let us notice that the magnetic charge $b = \frac{1}{2\pi}\int_{M^2}B\ d\sigma$
(which should be integer because of the quantisation condition) changes
under Laplace transformations according to the formula
$$
\tilde b = b + 2g-2
$$
as it follows from (\ref{102}) and the Gauss-Bonnet formula.
After $N$ steps we have $b_N = b_0 +2N(g-1)$.
Also from (\ref{*}) we have 
$$
\int_{M^2}(U_k+B_k)\ d\sigma = \int_{M^2}(U_{k-1} - B_{k-1})\ d\sigma
= \int_{M^2}(U_{k-1} + B_{k-1})\ d\sigma - 4\pi b_{k-1}.
$$
Since $U_0 + B_0 = 0$ 
it follows that 
$$\frac{1}{2\pi}\int_{M^2}(U_N+B_N)\ d\sigma = -\frac{1}{2\pi}\int_{M^2}c\ d\sigma
= -2Nb - N(N-1)(2g-2),$$ so we have the relation
\begin{equation}
c \frac{A(M^2)}{2\pi} = 2Nb_0 + N(N-1)(2g-2),
\label{rel}
\end{equation}
where $A(M^2) = \int_{M^2} \ d\sigma$ is the area of $M^2$.
For the sphere we have 
$c = Nb_0 - N(N-1)$, for the torus  $c A(M^2) = 4\pi Nb_0$,
for a surface of genus $g>1$ the relation $c (2g-2) = 2Nb_0 + N(N-1)(2g-2)$.

This determines the constant $c$ in the quasi-cyclic chain if we know the magnetic charge $b_0$
of the first operator and therefore imposes a quantisation condition on $c$ since $b_0$ is an integer.
In terms of this integer $b_0$ the sufficient conditions for the equations (\ref{A}) and (\ref{B})
to have a solution have the form (see the section 3)
$$
b_0 > g-1, b_0 + 2N(g-1) > g-1.
$$
For the topological sphere $S^2$ this is equivalent to the inequality $b_0 > 2N - 1$,
for the torus this simply means that $b_0$ is positive, for a surface of
genus $g>1$ we have just the first inequality $b_0 > g-1$.

\begin{theorem}

Let $U_k, B_k, k = 0,1,...,N$ satisfy the quasi-cyclic chain (\ref{*})
on a curved suface $M^2$ and let the magnetic charge of the first operator $b_0$
satisfy the conditions described above.
Then the last operator of the chain $L_N$ has two known energy levels
independent of the Aharonov-Bohm fluxes: the ground state
$\lambda = -c$ and $\lambda = 0$. The corresponding eigenfunctions can be found from the 
solutions of the first order equations (\ref{A}) and (\ref{B}) 
and for a large $b_0$ 
form the spaces of the dimensions $b_0 + (2N-1)(g-1)$ for $\lambda = -c$ and $b_0 - g +1$
for $\lambda = 0$ respectively.

\end{theorem}

{\bf Example 1.}  Quasi-cyclic chains of length $N=1$:
$$
\begin{array}{c}
U_0+B_0=0 \\
\ \\
U_1+B_1=U_0-B_0=-c
\end{array}
$$
implies that $U_0=-B_0=-c/2$ are constants, so that $L_0$ is the Landau operator on 
a curved surface $M^2$. 
Notice that 
$$
B_1 = B_0+\frac{1}{2} \Delta_h(\ln (-c)) -K=B_0-K,
$$
$$
U_1=U_0-2B_0+K=K-3B_0,
$$
so that in this case Laplace transformation coincides with one step of the 
factorisation procedure (\ref{24}):
$(B_0, -B_0)\to (B_0-K, K-3B_0), ~~~ B_0=\frac{c}{2}.$
Thus in this case we claim that the Schr\"odinger operator with the magnetic field $B= \frac{c}{2} - K$
and the potential $U = K - \frac{3c}{2}$ has two lowest energy levels known: 
$\lambda = -c$ and $\lambda = 0$  provided the quantisation and positivity 
conditions for $c$ are satisfied.

{\bf Example 2.} Quasi-cyclic chains with $N=2:$
$$
\begin{array}{c}
U_0+B_0=0 \Leftrightarrow U_0=-B_0; \\
\ \\
B_1=B_0+\frac{1}{2}\Delta_h (\ln B_0)-K, ~~~ U_1=U_0-B_0-B_1=-2B_0-B_1; \\
\ \\
U_2+B_2=U_1-B_1=-2B_0-2B_1=-c.
\end{array}
$$
Thus we have the following relation for the magnetic field $B_0$:
$$
2B_0+\frac{1}{2}\Delta_h (\ln B_0)-K=\frac{c}{2}
$$
or
$$
\Delta_h (\ln B_0)=c+2K-4B_0
$$
which after introducing $\varphi=\ln B_0$ takes the form
\begin{equation}
\Delta_h \varphi=c+2K-4 \ e^{\varphi}.
\label{L}
\end{equation}
When $K=0$ this reduces to the equation from \cite{NV97}. 

As well as in the flat case any solution of (\ref{L}) 
determines a Schr\"odinger operator in magnetic field with two integrable levels.

{\bf Remark.} When $c=0$ equation (\ref{L}) reduces to 
$$
\Delta_h \varphi=2K-4 \ e^{\varphi}
$$
which is a natural analogue of the well-known Liouville equation for a curved surface.
It can be transformed into the standard Liouville equation 
$\tilde \varphi _{z\bar z}=-e^{\tilde \varphi}$
by a substitution $\varphi = \tilde \varphi +2\ln h.$
When $c \ne 0$ the equation (\ref{L}) is probably non-integrable already in the flat case.

\section{Acknowledgments}

Our interest to the quantum problems with magnetic fields has been initiated by S.P.Novikov
to whom we would like to express a special gratitude.
We are grateful also to Nick Manton and John Samson for useful and stimulating discussions
and to A.Macfarlane, R.Seiler and F.Williams for sending us the copies of their papers
which appeared to be very helpful.

This research has been partially supported by EPSRC (grant No GR/M69548).

\end{document}